\documentclass[12pt]{scrartcl} 
\usepackage{hyperref}
\usepackage{amsmath}
\usepackage{todonotes}
\usepackage{amsfonts}
\usepackage{booktabs}
\usepackage{subfig}
\usepackage{apacite}
\usepackage{hyperref}

\providecommand{\keywords}[1]{\textbf{\textit{Keywords---}} #1}

\title{Gaussian Process Priors for Dynamic Paired Comparison Modelling}
\author{Martin Ingram* \\
		University of Melbourne, Australia \\
		\small email: \href{mailto:ingramm@student.unimelb.edu.au}{ingramm@student.unimelb.edu.au}}

\begin{document}
\maketitle

\abstract{
	Dynamic paired comparison models, such as Elo and Glicko, are frequently used for sports prediction and ranking players or teams. We present an alternative dynamic paired comparison model which uses a Gaussian Process (GP) as a prior for the time dynamics rather than the Markovian dynamics usually assumed. In addition, we show that the GP model can easily incorporate covariates. We derive an efficient approximate Bayesian inference procedure based on the Laplace Approximation and sparse linear algebra. We select hyperparameters by maximising their marginal likelihood using Bayesian Optimisation, comparing the results against random search. Finally, we fit and evaluate the model on the 2018 season of ATP tennis matches, where it performs competitively, outperforming Elo and Glicko on log loss, particularly when surface covariates are included.
}

\vspace{10px}

\vspace{10px}

*: Corresponding author

\newpage

\section{Introduction}

\subsection{Background and motivation}

As the name suggests, paired comparison models aim to predict the outcome of a comparison of two items. Although they are used for other purposes, such as in marketing research (for an overview, see for example \cite{bradley198414}), the focus of this paper is on the prediction of sporting outcomes. Here, the items being compared are players, and the outcome of the comparison depends on how much greater one player's skill is compared to the other's.

Perhaps the most widely used paired comparison model is the ``Bradley-Terry'' model \cite{bradley1952rank}. In this model, the likelihood can be written as:
\begin{align}
	  P(\textrm{y=1}|\theta_i, \theta_j) = \frac{1}{1 +
      e^{\theta_j - \theta_i}} = \textrm{logit}^{-1}(\theta_i - \theta_j)
      \label{eq:bradley-terry}
\end{align}
Here, $\theta_i$ and $\theta_j$ are latent abilities of player $i$ and $j$, which are assumed to be scalar, and $y=1$ is the indicator that player $i$ won the comparison, or match (a loss has $y=0$). Extensions to the Bradley-Terry model exist to model other outcomes such as draws \cite{rao1967ties}, but we will consider only the win/loss case in this paper.

In sports, players' abilities are likely vary over time. In tennis, a player may decline in performance due to an injury, or improve due to better training; in team sports, teams may improve their tactics, or gain and lose players to injury over time. Rather than keeping the latent abilities $\theta$ fixed, \emph{dynamic} paired comparison models allow them to vary over time. Here we consider two: the popular Elo model \cite{elo1978rating}, and an approximate Bayesian model known as Glicko \cite{glickman1999parameter}.

Elo was originally developed by Arpad Elo for ranking chess players. Its likelihood is a rescaled version of the Bradley-Terry likelihood:
\begin{align}
	P(\textrm{y=1}|\theta_i, \theta_j) = \frac{1}{1 + 10^{(\theta_j - \theta_i) / 400}}
	\label{eq:elo-likelihood}
\end{align}

Elo is an algorithm rather than a full probabilistic model. Players start with a rating $\theta_0 = 1500$. After each match, ratings are updated according to a simple formula:
\begin{align}
	\theta_i' = \theta_i + k \times (y - P(y|\theta_i, \theta_j))
\end{align}
More surprising results will produce larger updates, with the maximum update given by $k$. This update size $k$ can be chosen using, for example, by maximising the likelihood using Equation \ref{eq:elo-likelihood}.

Elo is a popular model in many sports. In tennis, a slight variation proposed by the website FiveThirtyEight \cite{morris_bialik_2015}, which uses Elo as a component in  has been shown to perform strongly, outperforming ten other published models for tennis prediction in a recent review \cite{kovalchik2016searching}. Elo models have also been shown to be accurate at predicting other sports, such as association football \cite{lasek2013predictive}.

While Elo is an algorithm rather than a probabilistic model, the Glicko model is derived as an approximation to a Bayesian dynamic paired comparison model. In this model, players are given an initial prior described by a univariate normal distribution. Glicko breaks time into periods, during which skills are assumed to be constant. Over time, these skills change according to a Markovian random walk, the next period a small normal jump away from the current period. The likelihood employed is the same as in Elo (Equation \ref{eq:elo-likelihood}). By approximating the likelihoods as one-dimensional Gaussians, the Glicko model is able to use  Kalman filter equations to recursively update player skills. For certain assumptions, Glicko recovers Elo as a special case \cite{glickman1999parameter}, and may be preferable if uncertainty information is of interest.

Elo and Glicko are popular and accurate models of player skill over time, but they have a number of shortcomings which we aim to address in this paper. Firstly, Elo and Glicko both assume that given the most recent rating, updates are conditionally independent of all previous ratings. However, it may be that teams or players follow trends, and a downward correction in the previous period could make a second decline more likely, for example. Secondly, Glicko and Elo have no obvious way to incorporate covariates which may help prediction. For instance, in tennis, players' skills are known to vary by playing surface, with some players performing better on one surface, such as clay, versus others (see for example \cite{mchale2011bradley}). While ad-hoc modifications can be made, such as fitting Elo models to each surface separately and then using a weighted sum to predict, a fully model-based approach would be preferable.

In this paper, we present a way to use a Gaussian Process models (GPs) \cite{rasmussen2004gaussian} for paired comparison modelling. Through the use of different kernel functions, the GP allows the exploration of different latent dynamics which are not necessarily Markovian. In addition, the GP formulation allows covariates to be incorporated naturally. We derive an approximate inference scheme based on the Laplace approximation and sparse linear algebra which, while more computationally intensive than Elo and Glicko, is still able to fit models to thousands of matches in a matter of seconds. We evaluate the model on the 2018 season of ATP tennis matches and show that it has lower log loss than Elo and Glicko, particularly when incorporating surface covariates. We believe that the model could be an interesting alternative to Elo and Glicko for dynamic paired comparison modelling.

\subsection{Other related work}

In addition to Elo and Glicko, there are some other related papers we would like to mention. This is not the first work that applies GPs to paired comparison modelling. In \cite{chu2005preference}, the authors present a framework to use GPs for preference learning. Although their context is different to that presented here (relating for example to housing choices, rather than picking a winner in a sporting contest), this problem is related to the paired comparison problem we consider here. Another related work is ``The Player Kernel'' \cite{DBLP:journals/corr/MaystreKFG16}, in which the authors use a GP together with a novel kernel function to predict outcomes in association football. 

Our contributions in this paper are to explore the use the GP as a dynamic rather than a static paired comparison model, as well as using sparse linear algebra to accelerate model fit.

\section{Methods}

\subsection{Gaussian Process Prior}
\label{sec:gp-prior}

\paragraph{Notation} To distinguish vectors from scalars more easily and to be more consistent with the notation commonly used in the GP literature, we switch from the Greek letter $\theta$ to the Roman letter $f$ for denoting player skill. Vectors will be denoted in lower-case bold face (e.g. $\textbf{f}$), matrices in upper-case bold face (e.g. $\textbf{K}$), and scalars in lower-case regular face (e.g. $f_{ij}$). Multivariate normal distributions with mean $\pmb{\mu}$ and covariance matrix $\pmb{\Sigma}$ on a random vector $\mathbf{x}$ are denoted as $\mathcal{N}(\mathbf{x} | \pmb{\mu}, \pmb{\Sigma})$, and univariate normal distributions with mean $\mu$ and variance $\sigma^2$ are written as $\mathcal{N}(x | \mu, \sigma^2)$.

To understand the Gaussian Process prior, we first discuss the Glicko model. In Glicko, the prior on each player's skill follows a random walk over time.  Collecting player $i$'s skills into the vector $\mathbf{f_i}$, the Glicko prior is:
\begin{align}
	P(\mathbf{f_i}) = \mathcal{N}(f_{i, 1}|\mu_0, \sigma_0^2)
 \prod_{t=2}^{n_p}\mathcal{N}(f_{i, t} | f_{i,t-1}, \eta^2) 
 \label{eq:randomwalkglicko}
\end{align}
Here, the first term represents the prior distribution for the first period, and the second encodes the random walk dynamics, with the time index $t$ running from 2 until $n_p$, the total number of periods. Each player's prior is independent, leading to the following joint prior:
\begin{align}
	P(\mathbf{f}) = \prod_{i=1}^{n_p}P(\mathbf{f_i})
	\label{eq:independence}
\end{align}
where $n_p$ is the number of players, and $\mathbf{f} = (\mathbf{f^{T}_1},...,\mathbf{f^{T}_{n_p}})^T$ is the concatenation of all player vectors $\mathbf{f_i}$.

In the Gaussian Process model, we also assume that each player's prior is independent. However, instead of the random walk prior, we place the following prior on $\mathbf{f_i}$:
\begin{align}
	P(\mathbf{f_i}) = \mathcal{N}(\mathbf{f_i} | \mathbf{0}, \mathbf{K_i})
	\label{eq:gp-prior}
\end{align}
In other words, we place a multivariate normal prior on the skill vector $\mathbf{f_i}$ with mean $\mathbf{0}$ and covariance matrix $\mathbf{K_i}$. The entries of this covariance matrix are obtained by evaluating a kernel function $k(t, t')$ for each pair of time points. For example, the entry $j, k$ is given by:
\begin{align}
	[\mathbf{K_{i}}]_{jk} = k(t_j, t_k)
\end{align}
One popular kernel function is the radial basis function kernel (RBF):
\begin{align}
	k(t, t') = \alpha^2 \textrm{exp}\left(-\frac{(t - t')^2}{2 \rho^2}\right)
\end{align}
Here, $\alpha^2$ governs the overall variance of the function and $\rho$ determines how quickly the covariance between different points decays with time: a large value of $\rho$ will result in the covariance falling off slowly leading to even distant points being positively correlated, whereas a small value will lead to $k(t, t')$ dropping to near zero even for small $(t - t')$, resulting in points being almost independent.

Several points are worth noting. Firstly, while the Glicko prior in Equation \ref{eq:randomwalkglicko} divides time into periods, this is not required for the GP prior. Instead, $\mathbf{f_i}$ has $n_i$ elements, with $n_i$ being the number of matches played by player $i$, and $\mathbf{K_i}$ is calculated using the kernel function evaluated using the time difference in days between matches.\footnote{In practice, we divide the time difference by 300 to put them on a smaller scale.}

Secondly, the kernel function can easily be extended to accept vectorial inputs. For example, the RBF kernel with so-called automatic relevance determination (ARD) can be written as follows:
\begin{align}
  k(\mathbf{x}, \mathbf{x'} | \alpha, \pmb{\rho}) =
  \alpha^2 \ \textrm{exp}\left(-\sum_{k=1}^{n_c}\frac{(x_k - x'_k)^2}
  {2\rho_k^2}\right)
  \label{eq:ard-rbf}
\end{align}
where $n_c$ is the number of dimensions of $\mathbf{x}$. We will use this later to incorporate covariates other than time into the model.

Finally, we would like to emphasise that in contrast to the random walk in Equation \ref{eq:randomwalkglicko}, the GP prior in Equation \ref{eq:gp-prior} does not enforce conditional independence assumptions about players' skill development over time. In the next section, we illustrate how the choice of different kernel functions leads to different assumptions about the evolution of player skills through time.

\subsection{Gaussian Processes as priors over functions}
\label{sec:gp-priors}

A useful way to think about Gaussian Processes is as priors over functions. In this section, we illustrate how the choice of kernel function affects this prior.

\begin{figure}
	\includegraphics[width=\textwidth]{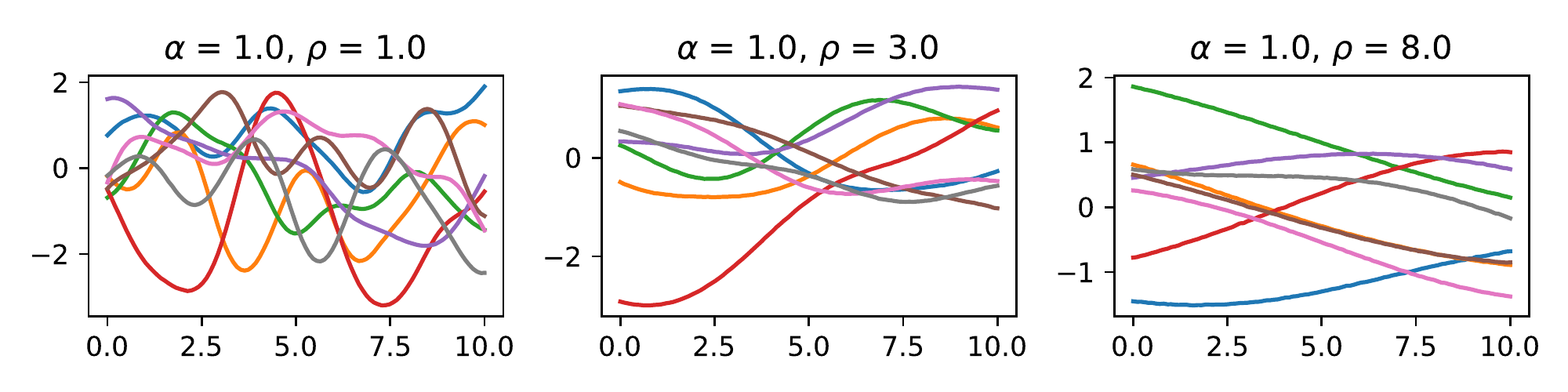}
	\caption{Draws from an RBF prior with the variance $\alpha^2$ fixed, and varying lengthscale $\rho$. As the lengthscale increases, the function varies more slowly as a function of $x$. The variance $\alpha^2$ governs the range of values in the $y$ direction; here, they range roughly from -2 and 2, as expected for a normal distribution with variance 1.}
	\label{fig:lengthscale-rbf}
\end{figure}

Figure \ref{fig:lengthscale-rbf} shows how the choice of lengthscale affects the draws from a GP with an RBF prior. As mentioned previously, shorter lengthscales lead to functions that vary more quickly, since the covariance between points drops off more quickly.

\begin{figure}
	\includegraphics[width=\textwidth]{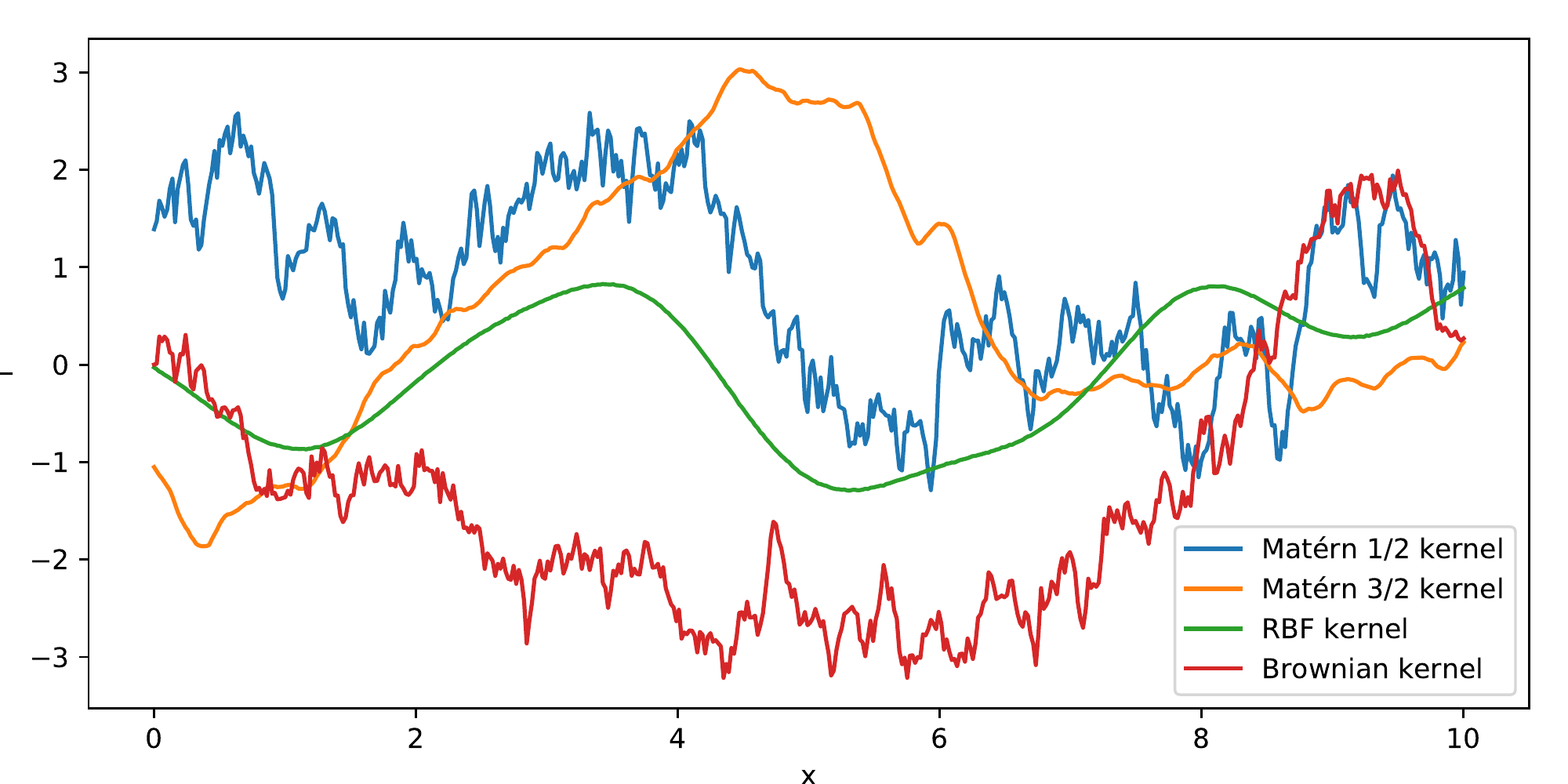}
	\caption{Draws of different kernels for the same variance (1) and lengthscale (also 1). Despite similar parameters, the functions differ significantly in their appearance. The RBF kernel appears smoothest, followed by the Mat\'ern 3/2 kernel. The Brownian and Mat\'ern 1/2 kernels are very jagged.}
	\label{fig:different-kernels}
\end{figure}

So far, we have discussed only the RBF kernel. However, this is only one choice among many possible kernels. Figure \ref{fig:different-kernels} shows samples drawn from priors defined by some other kernel functions. The draws illustrate that while some kernels put prior weight on very smooth functions (particularly the RBF kernel), others instead look very jagged (Mat\'ern 1/2 or Brownian kernel).

Kernel functions can also be combined. The two most common ways of combining kernels are addition and multiplication. Adding two kernel functions can be thought of as assuming that the function is a sum of two underlying functions with different properties (such as one fast-varying component and a slowly-varying one), while multiplication is particularly useful when combining kernels placed on different covariates. For example, we later multiply one kernel function placed on time with another placed on covariates, which leads to a covariance matrix that has its highest values for pairs of points that are both close in time and in terms of the covariates, and which takes on small values when either kernel function is small.

\subsection{Model conditional on kernel hyperparameters}

In this section, we assume that a kernel function and its hyperparameters (lengthscale and variance) have been chosen (we will discuss how to choose these parameters in the next section), and focus on inference for the parameters $\mathbf{f}$ given this kernel.

\paragraph{Full Prior} As discussed in section \ref{sec:gp-prior}, we place independent multivariate normal priors on each player's skill vector $\mathbf{f_i}$ (Equation \ref{eq:gp-prior}). Concatenating all these skill vectors into one long skill vector $\mathbf{f}$, and collecting the hyperparameters of the kernel into the vector $\pmb{\theta}$, leads to another joint multivariate normal prior:
\begin{align}
	P(\mathbf{f}|\pmb{\theta}) = \mathcal{N}(\mathbf{f} | \mathbf{0}, \mathbf{K})
\end{align}
Because of the assumption that each player's skill is independent of each other player's skill (Equation \ref{eq:independence}), this kernel matrix $\mathbf{K}$ is block diagonal, with each block given by each player's kernel matrix $\mathbf{K_i}$.

\paragraph{Likelihood} We assume that we have a dataset of $n$ matches. The likelihood for match $i$ is the Bradley-Terry likelihood:
\begin{align}
	P(y_i|f_{w(i)}, f_{l(i)}) = \textrm{logit}^{-1}(f_{w(i)} - f_{l(i)})
\end{align}
Here, $w(i)$ maps the match $i$ to the index of the match winner in the concatenated vector $\mathbf{f}$, and $l(i)$ maps to the loser. Since each match involves two players, the full vector $\mathbf{f}$ has $2n$ elements -- two entries for each match. The full likelihood factors over matches, so that:
\begin{align}
	P(\mathbf{y}|\mathbf{f}) = \prod_{i=1}^{n}\textrm{logit}^{-1}(f_{w(i)} - f_{l(i)})
	\label{eq:gp-lik}
\end{align}
Note that conditional on $\mathbf{f}$, the likelihood is independent of the kernel hyperparameters $\pmb{\theta}$.

\paragraph{Posterior} The posterior $P(\mathbf{f}|\mathbf{y}, \pmb{\theta})$ is given by Bayes' rule:
\begin{align}
	P(\mathbf{f}|\mathbf{y}, \pmb{\theta}) = \frac{P(\mathbf{f}|\pmb{\theta})P(\mathbf{y}|\mathbf{f})}{P(\mathbf{y}|\pmb{\theta})}
	\label{eq:posterior}
\end{align}

\subsection{Approximate inference given kernel hyperparameters}

The non-Gaussian Bradley-Terry likelihood in Equation \ref{eq:gp-lik} is not conjugate to the Gaussian prior, which means that there is no closed-form expression for the posterior in Equation \ref{eq:posterior}. Instead, we resort to the Laplace approximation to compute an approximate posterior \cite{rasmussen2004gaussian}.

The Laplace approximation first finds the minimum of the negative log posterior and then uses a Taylor expansion to approximate the function with a multivariate normal distribution. To find the mode, we initialise each element of $\mathbf{f}$ to zero and proceed using Newton's method until convergence:
\begin{align}
	\mathbf{f'} = \mathbf{f} - \mathbf{H}^{-1}\mathbf{j}
	\label{eq:newton}
\end{align}
Here, $\mathbf{H}^{-1}$ is the inverse of the Hessian of the negative log posterior evaluated at $\mathbf{f}$, and $\mathbf{j}$ is its Jacobian.

Calculating the update in Equation \ref{eq:newton} initially seems prohibitively expensive: since $\mathbf{f}$ has dimension $2n$, twice the number of matches, the Hessian can become very large. In tennis, roughly 2000 matches are played on the men's professional ATP tour every year. Calculating $\mathbf{H}^{-1}\mathbf{j}$ would thus seem to involve $\mathcal{O}((2n)^3)$ operations. However, we will show that the Hessian of the negative log posterior is in fact very sparse, which allows faster computation.

The negative log posterior, as a function of $\mathbf{f}$, is:
\begin{align}
	-\textrm{log}(P(\mathbf{f}|\mathbf{y}, \pmb{\theta})) &= -\textrm{log}P(\mathbf{f}|\pmb{\theta}) - \textrm{log}P(\mathbf{y}|\mathbf{f}) \\
	&= \frac{1}{2}\mathbf{f}^T\mathbf{K}^{-1}\mathbf{f} - \textrm{log}P(\mathbf{y}|\mathbf{f})
\end{align}
where we have dropped terms that are constant as a function of $\mathbf{f}$. The Jacobian of this quantity is:
\begin{align}
	-\nabla\textrm{log}(P(\mathbf{f}|\mathbf{y}, \pmb{\theta})) &= 
	\mathbf{K}^{-1}\mathbf{f} - \nabla\textrm{log}P(\mathbf{y}|\mathbf{f}) = \mathbf{j}
\end{align}
And the Hessian is:
\begin{align}
	-\nabla\nabla\textrm{log}(P(\mathbf{f}|\mathbf{y}, \pmb{\theta})) &= 
	\mathbf{K}^{-1} - \nabla\nabla\textrm{log}P(\mathbf{y}|\mathbf{f}) = \mathbf{H}
\end{align}
Since $\mathbf{K}$ is block diagonal, $\mathbf{K^{-1}}$ is also block diagonal and can be obtained cheaply by separately inverting each player's covariance matrix.

So far, we have been following the derivation in \cite{rasmussen2004gaussian} exactly. However, while their derivation assumes that the Hessian of the log likelihood is diagonal, this is not the case for the Bradley-Terry likelihood. The log likelihood is:
\begin{align}
	\textrm{log}P(\mathbf{y}|\mathbf{f}) = \sum_{k=1}^{n}g(f_{w(k)}-f_{l(k)})
	\label{eq:log-lik}
\end{align}
where we have set $\textrm{log}(\textrm{logit}^{-1}(x)) = g(x)$ for notational convenience. The partial derivative of the log likelihood with respect to a single element $f_i$ is:
\begin{align}
	\frac{\partial\textrm{log}P(\mathbf{y}|\mathbf{f})}{\partial{}f_i}
	&= g'(f_{w(k)} - f_{l(k)}) \textrm{ if w(k) = i or} \\
	&= -g'(f_{w(k)} - f_{l(k)}) \textrm{ if l(k) = i}
\end{align}
In other words, only one term in the sum in Equation \ref{eq:log-lik} remains after differentiation, corresponding to one match, and the sign changes depending on whether the player corresponding to element $i$ was the winner or loser in the match.

In both cases, the second derivatives are 
\begin{align}
	\frac{\partial^2\textrm{log}P(\mathbf{y}|\mathbf{f})}{\partial{}f_i^2} = g''(f_{w(k)}- f_{l(k)}) \textrm{ where w(k) = i or l(k) = i }
\end{align}
and
\begin{align}
	\frac{\partial^2\textrm{log}P(\mathbf{y}|\mathbf{f})}{\partial{}f_if_j} &= -g''(f_{w(k)} - f_{l(k)}) \textrm{ if $w(k) \in i, j$ and $l(k) \in i, j$, $i \neq j$ } \\
	&= 0 \textrm{ otherwise.}
\end{align}
In plain English, even though the Hessian of the log likelihood is of size $2n \times 2n$, it only has two non-zero elements per row: one on the diagonal, and one where $i$ and $j$ correspond to the winner and loser of a match, respectively. This leads to $4n$ total non-zero entries.

Since $\mathbf{K}^{-1}$ is sparse, and $-\nabla\nabla\textrm{log}P(\mathbf{y}|\mathbf{f})$ is sparse, the Hessian $\mathbf{H}$ is sparse, too. In addition, $\mathbf{H}$ is positive definite (we do not prove this here), which has two advantages: firstly, the negative log posterior function is convex, which makes the Newton iterations converge very quickly; and secondly, we can use a sparse Cholesky decomposition to efficiently solve each Newton update step $\mathbf{H}^{-1}\mathbf{j}$. We call the \texttt{CHOLMOD} library \cite{chen2008algorithm} from python using the \texttt{scikit-sparse} library to accomplish this.

Once the posterior mode is found, the Laplace approximation to the posterior $P(\mathbf{f}|\mathbf{y}, \pmb{\theta})$ is given by:
\begin{align}
	Q(\mathbf{f}|\mathbf{y}, \pmb{\theta}) = \mathcal{N}(\mathbf{f}|\mathbf{\hat{f}}, \mathbf{H}^{-1})
	\label{eq:laplace}
\end{align}
where $\mathbf{\hat{f}}$ is the mode of the negative log posterior, and $\mathbf{H}$ is its Hessian, evaluated at $\mathbf{\hat{f}}$.

\paragraph{Prediction} Once again following \cite{rasmussen2004gaussian}, under the Laplace approximation, the predictive mean $f_*$ at a new input $\mathbf{x_*}$ is:
\begin{align}
	\mathbb{E}_q[f_*|\mathbf{X}, \mathbf{y}, \mathbf{x}_*] = \mathbf{k}_*^T \mathbf{K}^{-1} \mathbf{\hat{f}}
\end{align}
where $\mathbf{k_*}$ is the vector obtained by evaluating the kernel function between all ``training'' inputs $\mathbf{X}$ (a $2n \times 1$ matrix if only time is used, and $2n \times n_c$ in the general case when covariates are used) and the ``test'' input $\mathbf{x}_*$. The predictive variance at $f_*$ is:
\begin{align}
	\mathbb{V}_q[f_*|\mathbf{X},\mathbf{y},\mathbf{x_*}] = k(\mathbf{x_*}, \mathbf{x_*})
	- \mathbf{k_*}^T \mathbf{K}^{-1} \mathbf{k_*}
	+ \mathbf{k_*}^T \mathbf{K}^{-1} \mathbf{H}^{-1} \mathbf{K}^{-1} \mathbf{k_*} 
\end{align}

\subsection{Selecting hyperparameters using the approximate log marginal likelihood}

In the previous section, we derived a procedure to approximate the posterior \emph{given} the kernel hyperparameters $\pmb{\theta}$, such as lengthscale and variance. We still require a procedure to set these. As is common in the GP literature, we choose these by maximising the log marginal likelihood of the data given the hyperparameters, $P(\mathbf{y}|\pmb{\theta})$ \cite{rasmussen2004gaussian}.

The approximate log marginal likelihood under the Laplace approximation is given by:
\begin{align}
	\textrm{log}Q(\mathbf{y}|\pmb{\theta}) = \textrm{log}P(\mathbf{\hat{f}}|\pmb{\theta}) 
	+ \textrm{log}P(\mathbf{y}|\mathbf{\hat{f}}) 
	+ n \textrm{log}(2 \pi)
	- \frac{1}{2} \textrm{log}|\mathbf{H}|
\end{align}
where $n$ is the number of matches as before. The first two terms are the unnormalised log posterior and the last term can once again be calculated efficiently using the sparse Cholesky decomposition of the matrix $\mathbf{H}$. In practice, we drop the $n\textrm{log}(2\pi)$ term since it does not depend on the model.

Ideally, we would derive the gradients of this log marginal likelihood and maximise it using these. However, the gradients involve third derivatives of the log likelihood, which are relatively straightforward to compute when the Hessian of the log likelihood is diagonal, but more difficult to derive for the Bradley-Terry likelihood.

The number of hyperparameters is small, however: the models we consider later have at most 8 hyperparameters. We thus maximise the log marginal likelihood using Bayesian Optimisation, using the python package \texttt{GPyOpt} \cite{gonzalez2016gpyopt}. Given a function and search bounds for its parameters, Bayesian Optimisation maximises the function by exploring the search space. It does this by fitting surrogate GP models to the function evaluations, trading off exploitation (searching near the current best set of parameters) with exploration (reducing overall uncertainty about the function). 

We compare the results obtained by Bayesian Optimisation with randomly exploring the search space. To perform the random search, we uniformly sample from the bounds of each parameter, evaluate the function at the point sampled, and repeat this procedure, recording the best parameter values and function value found.

\subsection{Experiments}

\subsubsection{Validation dataset and training procedure}

We use the 2018 season of the ATP, the men's professional tennis circuit, to evaluate the model's performance. The data was obtained using the \texttt{OnCourt} software\footnote{\url{http://www.oncourt.info/download.html}}. We discard the unusual Davis Cup and ATP Next Gen Finals tournaments, as well as limiting the dataset to the tour level, removing lower-tier Challenger and Qualification events. This resulted in a validation dataset of 2,623 matches. The majority of matches were played on hard courts (1,072), followed by clay (810), indoor hard (417) and grass courts (324).

We fit the GP models beginning in 2016 and predict each match in an iterative fashion: to predict the first day of matches, we fit the model using all data up to that day. After predicting, we include the match results of this day in the training set to predict the next day, and so on, until all matches have been predicted.

\subsubsection{Kernel choices}

The framework derived in this paper is very general, and it is not obvious which choice of kernel will perform best for tennis. Here, we experiment with three different choices:

\paragraph{Experiment 1} In the first experiment, we fit a single Mat\'ern 3/2 kernel to the data. This kernel has two free parameters: the standard deviation and lengthscale. We set the search bounds of the standard deviation to be (0.01, 2), and (0.1, 10) for the lengthscale, corresponding to a lengthscale between 30 and 3,000 days.

\paragraph{Experiment 2} We exploit the fact that two kernel functions added together are also a valid kernel function by combining one Mat\'ern 1/2 kernel and one Mat\'ern 3/2 kernel. As Figure \ref{fig:different-kernels} shows, the Mat\'ern 3/2 kernel is relatively smooth, while the 1/2 kernel is quite jagged. We hypothesised that perhaps a tennis player's skill evolution could be modelled by a slow smooth component and a faster, more jagged component, hence this choice of kernels. The combined kernel has four free parameters -- lengthscale and standard deviation for each kernel -- which are given the same bounds as the single kernel in experiment 1.

\paragraph{Experiment 3} Finally, we investigate the utility of adding surface covariates to the model. We do this by one-hot encoding the surfaces into a $2n \times 4$ matrix and placing an RBF kernel with automatic relevance determination (see Equation \ref{eq:ard-rbf}) on this matrix. We then combine this kernel with a single Mat\'ern 3/2 kernel on the time dimension by multiplying the kernels together. The resulting kernel has six hyperparameters: two for the Mat\'ern kernel on time (lengthscale and standard deviation), and four lengthscales for the surfaces.

\subsubsection{Kernel hyperparameter fitting}

To fit the hyperparameters for each experiment, we maximise the log marginal likelihood using the 2016 and 2017 seasons. To investigate how many evaluations are required, we set the number of iterations in Bayesian Optimisation to a grid of 10, 50, 100 and 200, and run the optimisation 10 times for each. Each time, we record the optimal hyperparameters that were found as well as the best function value. To investigate the benefit of Bayesian Optimisation, we also run the same experiments using random search.

\subsubsection{Baseline models}

We choose Elo and Glicko as the baseline models. Elo has a single hyperparameter (the learning rate $k$); Glicko has two (the initial standard deviation and the period-to-period variance). We set these by optimising the log likelihood on the training set. For Glicko, a period length has to be chosen; we found that a period length of 1 performed best.

Previous work suggests that Elo performs better with earlier start dates \cite{kovalchik2016searching}. We thus also fit Elo and Glicko from the very start of the dataset (August 2002) to compare against the 2016 start.

\subsubsection{Evaluation metrics}

We compare the models using two metrics: log loss and accuracy. Log loss is the negative mean log likelihood:
\begin{align}
	\textrm{log loss}(\textbf{y}, \textbf{p}) = -\frac{1}{n} \sum_{i=1}^{n} \left[ 
	y_i \textrm{log}(p_i) + 
	(1 - y_i) \textrm{log}(1 - p_i) \right]
\end{align}
where $n$ is the number of matches in the evaluation set, $\textbf{y}$ is the vector of outcomes, and $\textbf{p}$ is the vector of probabilities predicted by the model. Log loss provides a good estimate of model calibration, penalising the model harshly for outcomes it considers unlikely.

Accuracy is simply the fraction of binary outcomes correctly predicted by the model:
\begin{align}
	\textrm{accuracy}(\textbf{y}, \textbf{p}) = \frac{1}{n} \sum_{i=1}^{n} \left[
	y_i \mathbb{I}(p_i > 0.5) + 
	(1 - y_i) \mathbb{I}(p_i \leq 0.5) \right]
\end{align}
where $\mathbb{I}$ is the indicator function and $\textbf{y}$ and $\textbf{p}$ are defined as before. We include accuracy as a metric because it is more easily interpretable than log loss.

\subsection{Speed test}

To investigate how well the model scales with more data, we record the time the model takes to fit datasets containing varying numbers of matches $n$, ranging from $n=1306$ (June 2017 - December 2017) to $n=\textrm{13,159}$ (January 2013 - December 2017). We run each set ten times and record the mean time taken in seconds. Each time, we record how long it takes to find the posterior mode $\hat{\mathbf{f}}$ plus the time taken to calculate the approximate log marginal likelihood. We ran this speed test on a 2017 MacBook Pro with a 3.1GHz Intel Core i5 and 16GB of RAM.

\section{Results}

\subsection{Speed}

\begin{figure}
\centering
\includegraphics[width=0.8\textwidth]{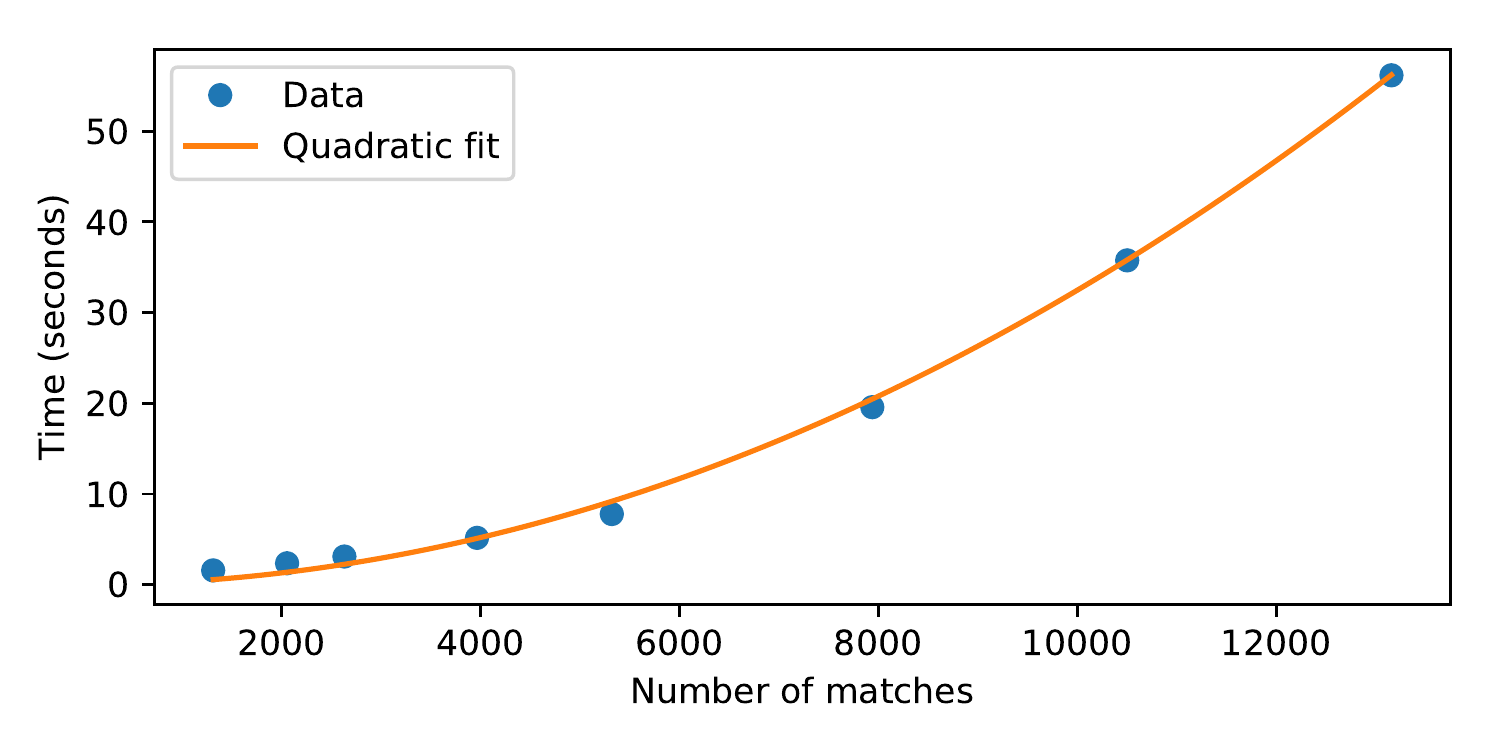}
\caption{Time required to fit datasets containing a varying number of matches. The scaling seems to be very well described by a quadratic fit of the form $3.25 \cdot 10^{-7} n^2$, shown in yellow, where $n$ is the number of matches.}
\label{fig:speed-scaling}
\end{figure}

Figure \ref{fig:speed-scaling} shows the time taken to find the mode $\mathbf{\hat{f}}$ of the posterior plus the calculation of the approximate log marginal likelihood for varying numbers of matches. The increase in time taken closely resembles the quadratic function $3.25 \cdot 10^{-7} n^2$, where $n$ is the number of matches. The dataset used to optimise the hyperparameters consisted of 5,323 matches, which took 7.8 seconds to fit. The largest dataset tested contained 13,159 matches and took 56.2 seconds.

\subsection{Hyperparameter optimisation}

\begin{figure}
\includegraphics[width=\textwidth]{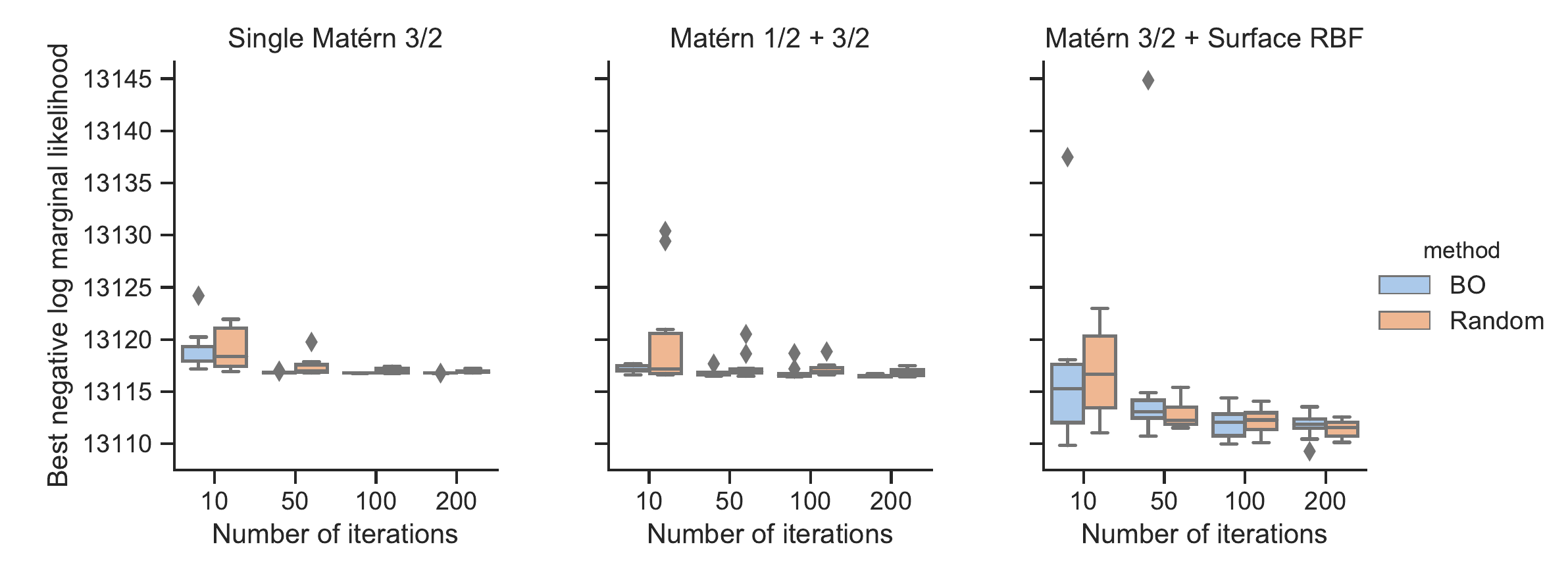}
\caption{Results from the hyperparameter optimisation. The y-axis shows the negative marginal log likelihood (lower indicates better fit); the x-axis the number of iterations used in the search. ``BO'' is short for Bayesian Optimisation; ``Random'' for random search. The single Mat\'ern kernel and combined kernels in panels 1 and 2 converge to very similar values. Adding the surface information appears to improve the fit (panel 3).}
\label{fig:experiment-results}
\end{figure}

Figure \ref{fig:experiment-results} shows the result of fitting the hyperparameters to the kernels in experiments (1) to (3) outlined previously. 

In Experiment (1), the single Mat\'ern 3/2 kernel, both approaches appear to converge to an optimum fairly quickly. After 10 iterations, there is some variation in the results found, but this decreases for 50 iterations, and for the runs searching for 100 and 200 iterations, random search and Bayesian Optimisation both find values close to 13116.8. The results for Bayesian Optimisation are slightly lower in general, and for 50 iterations, random search has a notably worse run returning a negative marginal log likelihood of around 13120, while all runs of Bayesian Optimisation return similar optima. The optimum lengthscale found was 5.29 (about 1,587 days), and the standard deviation was 0.882.

Experiment (2), the combined Mat\'ern 1/2 and 3/2 kernels, appears to result in a similar optimum as experiment (1): the best value found was 13116.4. This corresponded to setting the Mat\'ern 3/2 kernel to a lengthscale of 9.87 (2,961 days) with a standard deviation of 0.687, and the 1/2 kernel to a lengthscale of 7.95 (2,385 days) with standard deviation of 0.571. Both random search and Bayesian Optimisation return suboptimal outliers for 50 and 100 iterations in some of the 10 runs, but return similar results after 200 iterations and appear to have settled close to an optimal value.

 Experiment (3), combining a Mat\'ern 3/2 kernel on time with an ARD RBF kernel on surface, appears harder to optimise, with neither method honing in exactly on an optimum even after 200 iterations. The best value found was 13109.3, found by Bayesian Optimisation after 200 iterations, which is an outlier in the box plot. This is lower than the negative marginal log likelihood found for experiments (1) and (2). The optimum lengthscale for the Mat\'ern 3/2 kernel was 5.25 and the standard deviation 0.897, which is very similar to the kernel in experiment (1). The lengthscales for clay, grass, hard and indoor hard were 2.57, 2.31, 7.47 and 2.08, respectively. 
 
 We convert these lengthscales to the correlation matrix shown in Table \ref{tab:correlations} using Equation \ref{eq:ard-rbf}. Correlations range from 72\% (grass and indoor hard) to 81\% (hard and clay). Hard court results are estimated to be most correlated to other surfaces, with correlations ranging from 78\% (indoor hard) to 81\% (clay).

\begin{figure}
\centering
\begin{tabular}{lrrrr}
\toprule
{} &  clay &  grass &  hard &  indoor\_hard \\
\midrule
clay        &  1.00 &   0.75 &  0.81 &         0.73 \\
grass       &  0.75 &   1.00 &  0.80 &         0.72 \\
hard        &  0.81 &   0.80 &  1.00 &         0.78 \\
indoor\_hard &  0.73 &   0.72 &  0.78 &         1.00 \\
\bottomrule
\end{tabular}
\caption{Correlation matrix implied by the lengthscales found for the ARD RBF kernel. Performances on all surfaces are strongly correlated, with values ranging from 72\% (grass and indoor hard) to 81\% (hard and clay).}
\label{tab:correlations}
\end{figure}

\subsection{Evaluation on 2018 ATP season}

Table \ref{tab:eval-scores} shows the metrics of each model on the 2018 evaluation set. The log loss scores range from 0.631 (GP with surface covariates) to 0.639 (Elo starting in 2016). Accuracy scores range from 62.8\% (Elo and Glicko with a 2016 start) to 64.4\% (Elo with 2002 start).

Starting Elo and Glicko in 2002 improves their log loss (0.637 to 0.635 for Glicko, 0.639 to 0.638 for Elo) and accuracy (62.8\% for both to 64.1\% for Glicko and 64.4\% for Elo). Even with the 2002 start, the log loss is higher than for the GP models, but accuracy is slightly higher (64.4\% for Elo compared to 63.7\% for the best GP model).

\begin{figure}
\centering
\begin{tabular}{lrr}
\toprule
{} &  Log loss &  Accuracy \\
Model                        &           &           \\
\midrule
GP Mat\'ern 3/2 + Surface 2016 &     \textbf{0.631} &     0.636 \\
GP Mat\'ern 3/2 2016           &     0.634 &     0.637 \\
GP Mat\'ern 3/2 + 1/2 2016     &     0.634 &     0.634 \\
Glicko 2002                  &     0.635 &     0.641 \\
Glicko 2016                  &     0.637 &     0.628 \\
Elo 2002                     &     0.638 &     \textbf{0.644} \\
Elo 2016                     &     0.639 &     0.628 \\
\bottomrule
\end{tabular}
\caption{Evaluation metrics on the 2018 evaluation set. On log loss, the surface-specific GP performs best across all models (0.631) and the GP models outperform all Elo and Glicko models, even when fit starting in 2002. On accuracy, Elo and Glicko perform better when starting in 2002, with the Elo model showing the best accuracy (64.4\%). Among the models fit with a 2016 start, the Mat\'ern 3/2 model has the best accuracy (63.7\%).}
\label{tab:eval-scores}
\end{figure}

\subsection{Prediction example compared to Elo and Glicko}

\begin{figure}
	\includegraphics[width=\textwidth]{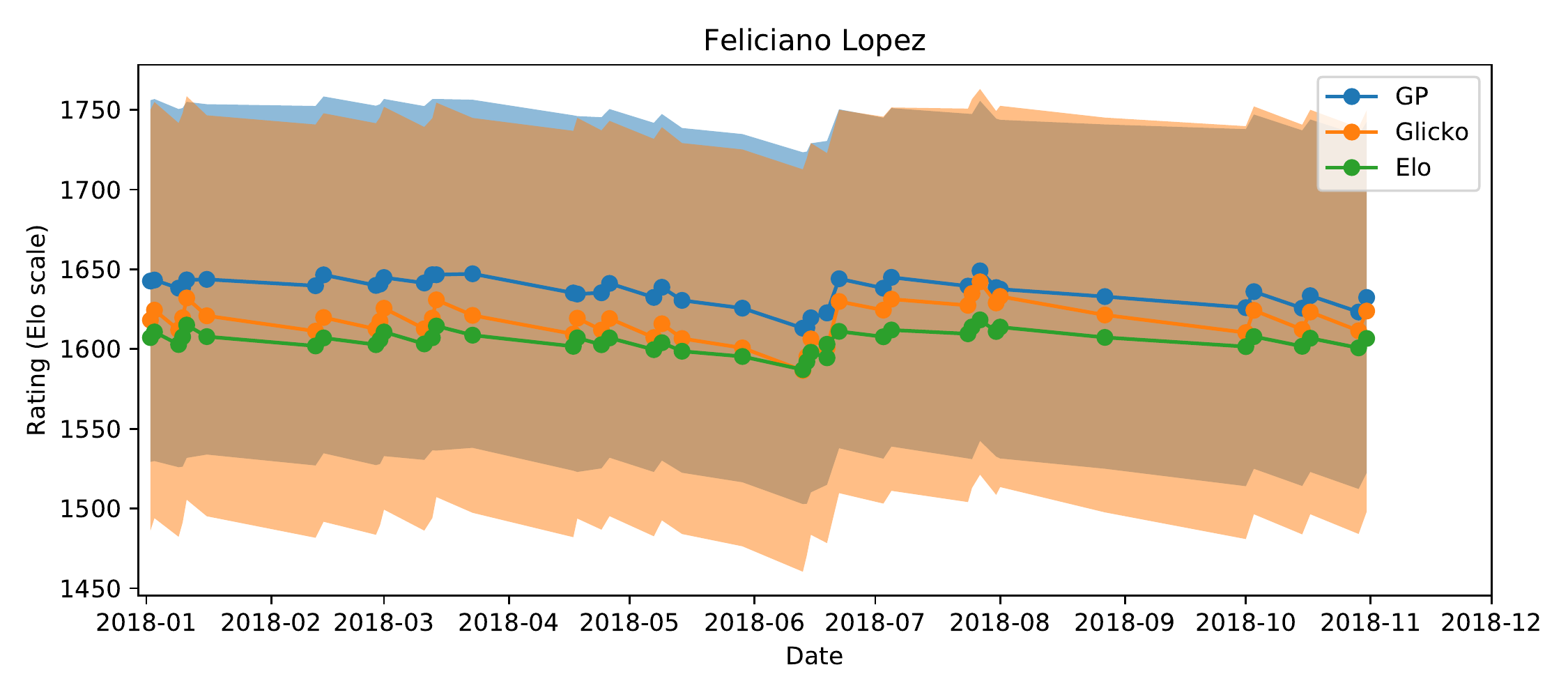}
	\caption{An example of the predictions made by the Mat\'ern 3/2 GP model on the 2018 evaluation set compared to Elo and Glicko. The shaded regions show credible intervals defined by two standard deviations from the approximate posterior for Glicko and the GP model (Elo does not provide uncertainty estimates).}
	\label{fig:feliciano-prediction}
\end{figure}

Figure \ref{fig:feliciano-prediction} compares the predictions made by the GP with the single Mat\'ern 3/2 kernel against Elo and Glicko on the 2018 season for an example player (Feliciano Lopez). We compare the predictions on the Elo scale, rather than the logit scale, by multiplying the logits by $\frac{400}{\textrm{log}(10)}$ and adding $1500$. This transformation can be straightforwardly derived from equations \ref{eq:bradley-terry} and \ref{eq:elo-likelihood}. 

Elo, Glicko and the GP model make similar adjustments to Lopez's rating over the course of the season. The credible interval is slightly narrower than that predicted by Glicko, and the GP rates Feliciano Lopez somewhat more highly than Glicko and Elo throughout the season.

\subsection{Illustration of covariates}

\begin{figure}
	\includegraphics[width=\textwidth]{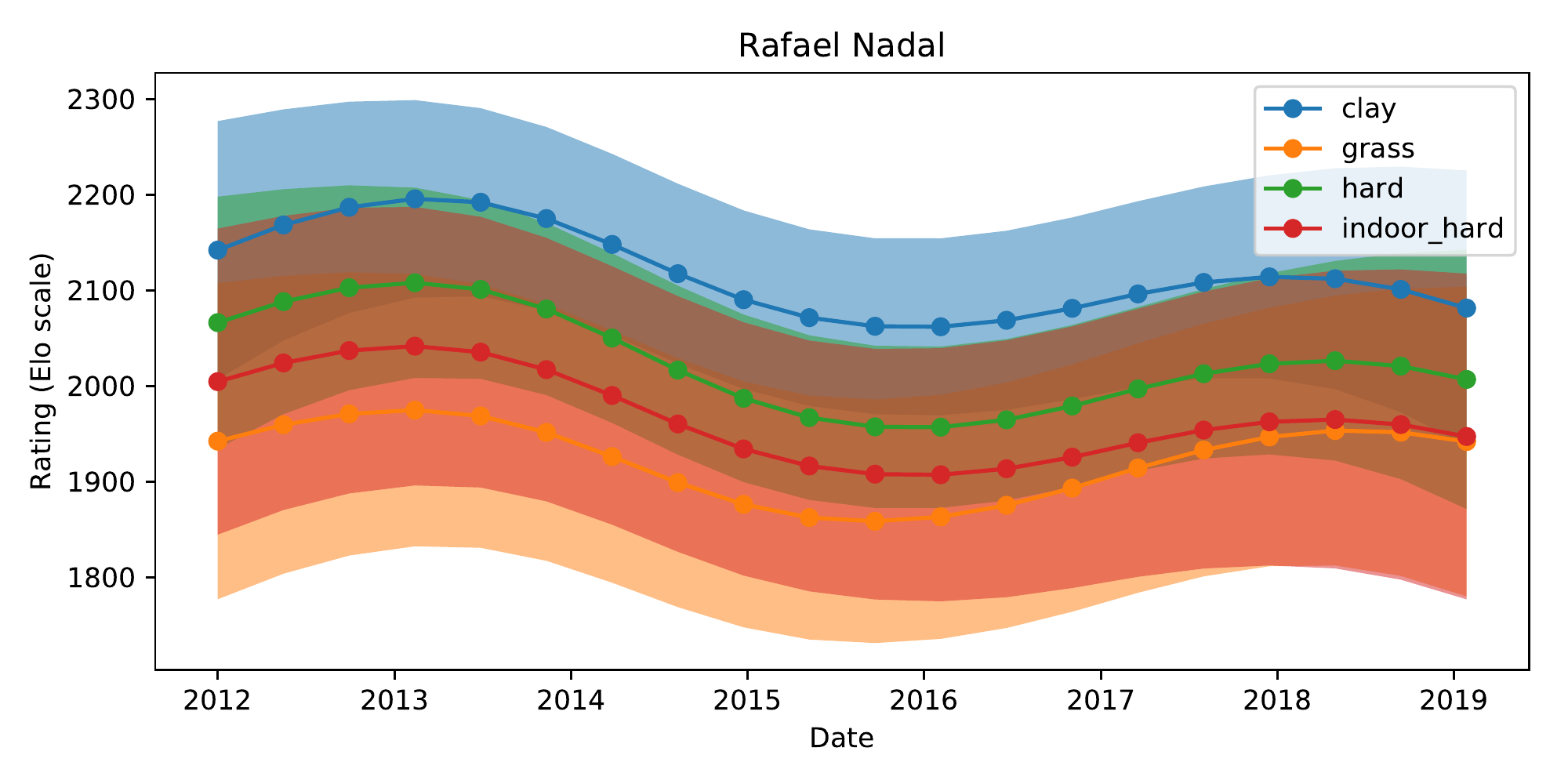}
	\caption{Rafael Nadal's ratings on different surfaces over time, as inferred by the surface-specific model. The bands show credible intervals corresponding to two standard deviations. The differences between surfaces are considerable, with Nadal's mean clay ability estimated to be around 2,200 points in 2013, compared to a grass rating of around 1,950. The gaps between skills are relatively constant, except the grass court rating, which appears to be rising in recent years.}
	\label{fig:nadal_surface}
\end{figure}

We fit the model from 2012 onwards to illustrate how the surface covariates affect the predictions of the latent ability. Figure \ref{fig:nadal_surface} shows an example of the surface abilities inferred for Rafael Nadal, one of the most successful players in recent years. The figure shows the posterior estimates of Nadal's skill with the model fit to the entire dataset. These estimates are notably smoother than those shown in Figure \ref{fig:feliciano-prediction}. This is because figure \ref{fig:feliciano-prediction} displays the predicted skill using only information leading up to each match, while Figure \ref{fig:nadal_surface} uses both information before and after the match.

The figure shows that differences between surfaces are large, with Nadal's rating highest on clay, followed by hard courts, indoor hard courts, and finally grass courts. Overall, the shape of the rating evolution is similar for all court types, but Nadal's grass court ability is estimated to have improved more rapidly since 2017 than his skill on indoor hard courts.

\begin{figure}
	\includegraphics[width=\textwidth]{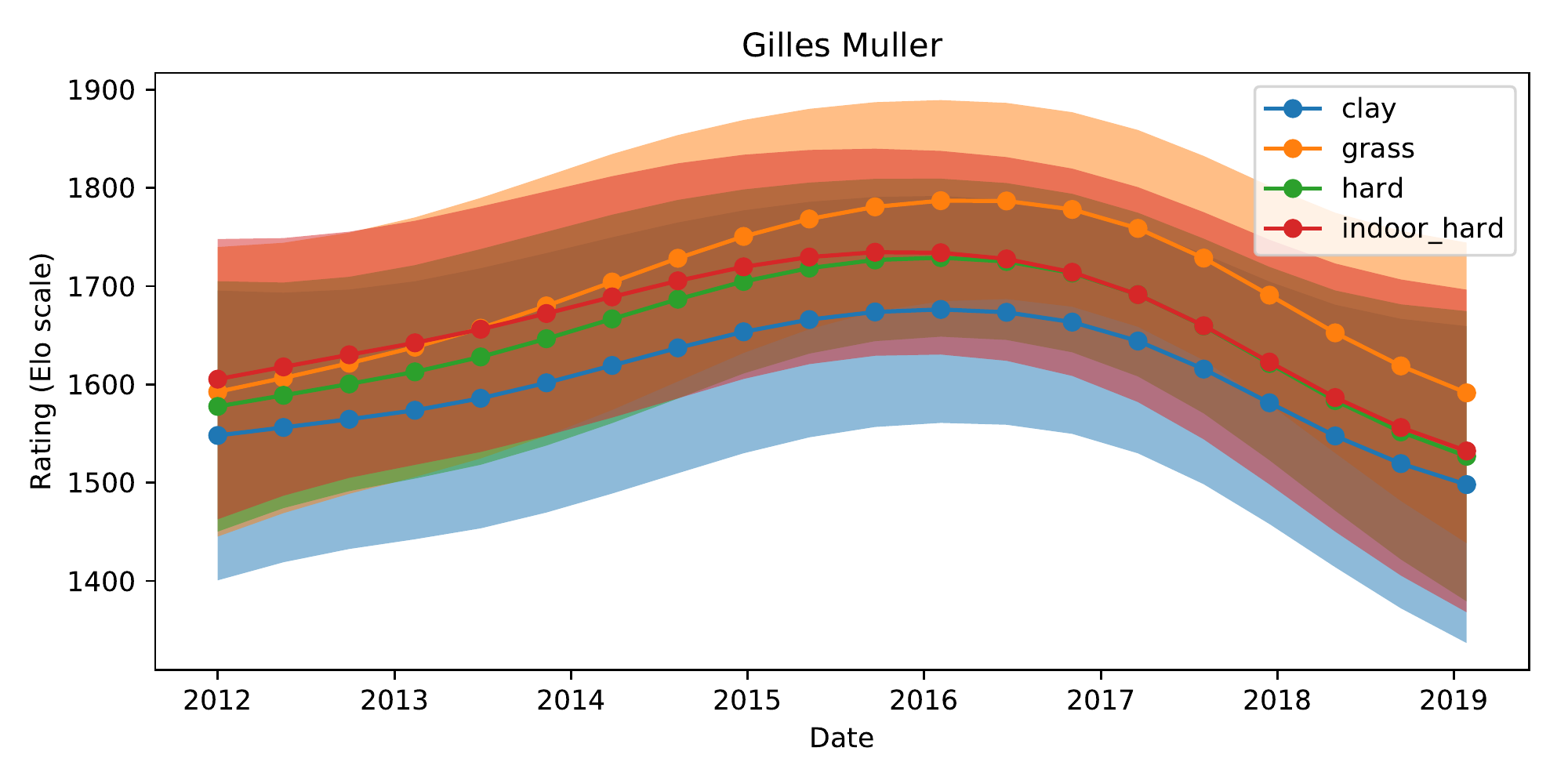}
	\caption{Gilles Muller's ratings on different surfaces over time, as inferred by the surface-specific model. As in Figure \ref{fig:nadal_surface}, the bands show credible intervals corresponding to two standard deviations.}
	\label{fig:muller_surface}
\end{figure}

Gilles Muller, shown in Figure \ref{fig:muller_surface}, is another striking example. In 2012, his skills on different surfaces were estimated to be similar, but over time, his grass court rating improved dramatically.

\subsection{Ranking example}

Table \ref{tab:ranking-2018} illustrates the model's ability to rank players. The rankings displayed were obtained by fitting the surface-specific GP model starting in 2016 up to the end of 2018 and predicting each player's rating at the end of the 2018 season (31st December 2018). Rafael Nadal, Novak Djokovic and Roger Federer are ranked highly on all four surfaces, with Nadal ranked first on three of the four. Nadal's ranking is highest on clay (2061), where he is ranked almost 150 points ahead of Novak Djokovic (1916), and lowest on indoor hard (1932), where Federer is ranked slightly ahead (1933). Standard deviations of the ratings appear to be highest on grass courts (72 to 83) and lowest on hard courts (58 to 70), which is likely a consequence of there being considerably fewer grass court matches in the dataset than hard court matches.

Table \ref{tab:ranking-2018} also highlights differing surface specialisations among the top players. Dominic Thiem is ranked sixth on clay courts but is ranked outside the top 8 on all other surfaces. Novak Djokovic, on the other hand, has relatively consistent ratings across all surfaces.

\begin{figure}
\centering
\subfloat[Clay]{

\begin{tabular}{lrr}
\toprule
{} &  mean &  sd \\
Player                &       &     \\
\midrule
Rafael Nadal          &  2061 &  73 \\
Novak Djokovic        &  1916 &  68 \\
Roger Federer         &  1914 &  74 \\
Alexander Zverev      &  1882 &  64 \\
Juan Martin Del Potro &  1848 &  70 \\
Dominic Thiem         &  1841 &  62 \\
Kei Nishikori         &  1808 &  66 \\
Marin Cilic           &  1793 &  67 \\
\bottomrule
\end{tabular}

}%
\subfloat[Grass]{
\begin{tabular}{lrr}
\toprule
{} &  mean &  sd \\
Player                &       &     \\
\midrule
Rafael Nadal          &  1956 &  83 \\
Novak Djokovic        &  1935 &  73 \\
Roger Federer         &  1925 &  74 \\
Juan Martin Del Potro &  1848 &  77 \\
Marin Cilic           &  1831 &  72 \\
Alexander Zverev      &  1814 &  73 \\
Kevin Anderson        &  1794 &  72 \\
Karen Khachanov       &  1780 &  73 \\
\bottomrule
\end{tabular}
}

\subfloat[Hard]{
\begin{tabular}{lrr}
\toprule
{} &  mean &  sd \\
Player                &       &     \\
\midrule
Rafael Nadal          &  1997 &  70 \\
Roger Federer         &  1958 &  66 \\
Novak Djokovic        &  1949 &  63 \\
Juan Martin Del Potro &  1893 &  59 \\
Alexander Zverev      &  1840 &  58 \\
Kevin Anderson        &  1808 &  58 \\
Kei Nishikori         &  1808 &  61 \\
Marin Cilic           &  1803 &  61 \\
\bottomrule
\end{tabular}
}%
\subfloat[Indoor hard] {
\begin{tabular}{lrr}
\toprule
{} &  mean &  sd \\
Player                &       &     \\
\midrule
Roger Federer         &  1933 &  74 \\
Rafael Nadal          &  1932 &  88 \\
Juan Martin Del Potro &  1898 &  76 \\
Novak Djokovic        &  1897 &  77 \\
Alexander Zverev      &  1822 &  69 \\
Kei Nishikori         &  1811 &  66 \\
Kevin Anderson        &  1789 &  67 \\
Marin Cilic           &  1783 &  72 \\
\bottomrule
\end{tabular}
}

\caption{Ratings at the end of the 2018 ATP season (31st December 2018), obtained from the surface-specific model with a fit starting in 2016. Each subtable shows the eight players with the highest means on each surface. For each player, their mean rating and its standard deviation (``sd'') is displayed.}

\label{tab:ranking-2018}

\end{figure}

\section{Discussion}

\subsection{Setting kernel hyperparameters}

In Figure \ref{fig:experiment-results}, we compared two strategies for selecting the kernel hyperparameters: random search and Bayesian Optimisation. Both seemed to consistently find an optimum in the case of the single Mat\'ern and two Mat\'ern kernels using 200 iterations, but the additional parameters involved in the surface kernel seemed challenging, with the best value being an outlier after 200 runs. This suggests that the 200 iterations may not have been sufficient to ensure convergence, and longer runs may be desirable.

In the comparison of random search and Bayesian Optimisation, Bayesian Optimisation tended to perform slightly better, finding the optimum more quickly for experiments (1) and (2), and performing similarly in experiment (3). We used the default arguments in \texttt{GPy}, which may not have been ideal: they consist of using point estimates to fit the hyperparameters of the surrogate model and use a particular acquisition function to choose where to evaluate the function next. Another inference method, such as Markov Chain Monte Carlo, as well as different acquisition functions, may perform better.

Overall, having to select the hyperparameters using gradient-free optimisation likely limits the number of parameters that can be fit within a reasonable length of time. Future work may include tackling the derivation of gradients, which would allow a much larger number of covariates to be fit.

\subsection{Choosing the best kernel}

As discussed in section \ref{sec:gp-priors}, the Gaussian Process framework allows modellers to experiment with a large variety of kernels to best fit their data. This freedom however can also make it difficult to choose which combination of kernels to use. For the tennis prediction example, we experimented with three different kernels, but other combinations may perform better. It would be interesting to attempt to adapt the automatic kernel discovery work in \cite{duvenaud2014automatic} to the GP presented here to investigate whether the automated procedure can find better models.

\subsection{Comparison against Elo and Glicko}

We believe that the model presented may be preferable to Elo and Glicko in certain situations. While it is more computationally expensive, the ability to fit thousands of matches in seconds should be sufficient for many modelling applications. It is also somewhat harder to implement, but we hope to mitigate this by providing code to fit the model online\footnote{\url{https://github.com/martiningram/paired-comparison-gp-laplace}}.

Aside from these drawbacks, the model has a number of advantages. It performs somewhat better on the evaluation dataset presented in the paper, particularly when adding surface covariates. Given that Elo outperformed other published prediction models in a previous review of tennis models \cite{kovalchik2016searching}, we believe this makes it quite a strong prediction model.

We also believe that the ability to combine kernels could be interesting to explore further. In tennis, this seemed to add little to model fit, with both kernels set to very long lengthscales (2,961 and 2,385 days) and similar values of the marginal log likelihood. However, this may be different in other sports.

Finally, the ability to include covariates sets the model apart from the baseline models. Adding surface covariates improved model fit considerably on the tennis dataset, and other covariates may further improve model fit. In other sports, other covariates may be of interest; for example, in chess, matches are sometimes played with different time limits, which may be analogous to the surface effect in tennis.

\section*{Acknowledgements}

This research was partially supported by the Melbourne Research Scholarship (MRS).

\bibliography{references.bib}
\bibliographystyle{apacite}

\end{document}